\documentclass[
  aps,
  prl,
  reprint,
  superscriptaddress,
  nofootinbib
]{revtex4-2}

\usepackage{amsmath,amssymb,bm}
\usepackage{graphicx}
\usepackage{hyperref}
\usepackage{xcolor}

\begin{document}

\title{Nonequilibrium thermodynamics of feedback-control: \\a phase-space perspective}

\author{Jos\'e A. Almanza-Marrero}
\email{joseantonio@ifisc.uib-csic.es}
\affiliation{Institute for Cross-Disciplinary Physics and Complex Systems IFISC (UIB-CSIC), Campus Universitat Illes Balears, E-07122 Palma de Mallorca, Spain}

\date{\today}

\begin{abstract}
Maxwell demons can convert knowledge into thermodynamic advantage by using measurement-acquired information. However, standard formulations of this problem assume that the demon has access to the system’s entire phase space, an assumption that fails in many practical applications. Here, I address this problem by deriving fluctuation relations for general feedback-controlled systems, including such singular cases. The central quantity is a generalized notion of unavailable information: a portion of the acquired information that cannot be used to extract useful work.
Finally, a Szilard engine with finite resolution illustrates these results. In the high-resolution limit, the information acquired by the demon may diverge, yet the extractable work remains finite. This work shows that this occurs because an equally divergent amount of information becomes unavailable.
\end{abstract}

\maketitle

\paragraph{Introduction.} 

Fluctuation relations provide exact constraints on nonequilibrium processes far from equilibrium. In their standard form, they relate the probability of a forward trajectory to its time reversal and imply integral identities such as the Jarzynski equality~\cite{Jarzynski97,Jarzynski97b} and the Crooks relation~\cite{Crooks99}. These identities impose strong constraints on the statistics of work and entropy production in driven systems and have been extensively reviewed in the literature~\cite{Seifert12,Jarzynski11,Campisi11} and have been also probed experimentally in several mesoscopic platforms, including colloidal particles, biomolecules, and electronic nanodevices~\cite{Liphardt02,Collin05,Toyabe10}.

The situation changes qualitatively when measurement and feedback are taken into account. This possibility was first envisioned by Maxwell, who imagined an intelligent being capable of sorting fast and slow molecules by observing them individually, producing an apparent violation of the second law~\cite{Maxwell71,szilard29}. The paradox was later resolved by recognizing that the erasure of the demon memory carry an intrinsic thermodynamic cost~\cite{Landauer61,Bennett82}, establishing that the acquisition and processing of information are themselves thermodynamic operations. This idea has been formalized within feedback-controlled stochastic thermodynamics, where the protocol driving a system depends on the outcome of measurements performed on it, and information becomes a thermodynamic resource~\cite{Horowitz10,SagawaUeda08,SagawaUeda10,SagawaUeda12,SagawaUeda12-2,Parrondo15}. This leads to generalized  relations in which entropy production is supplemented by information-theoretic terms, allowing negative average entropy production of the controlled system as long as the acquired information is properly accounted for. A central result in thermodynamics of information is the fluctuation theorem involving the stochastic information or transfer entropy acquired by the controller~\cite{Horowitz10,SagawaUeda12}   
\begin{figure}
    \centering
    \includegraphics[width=0.7\linewidth]{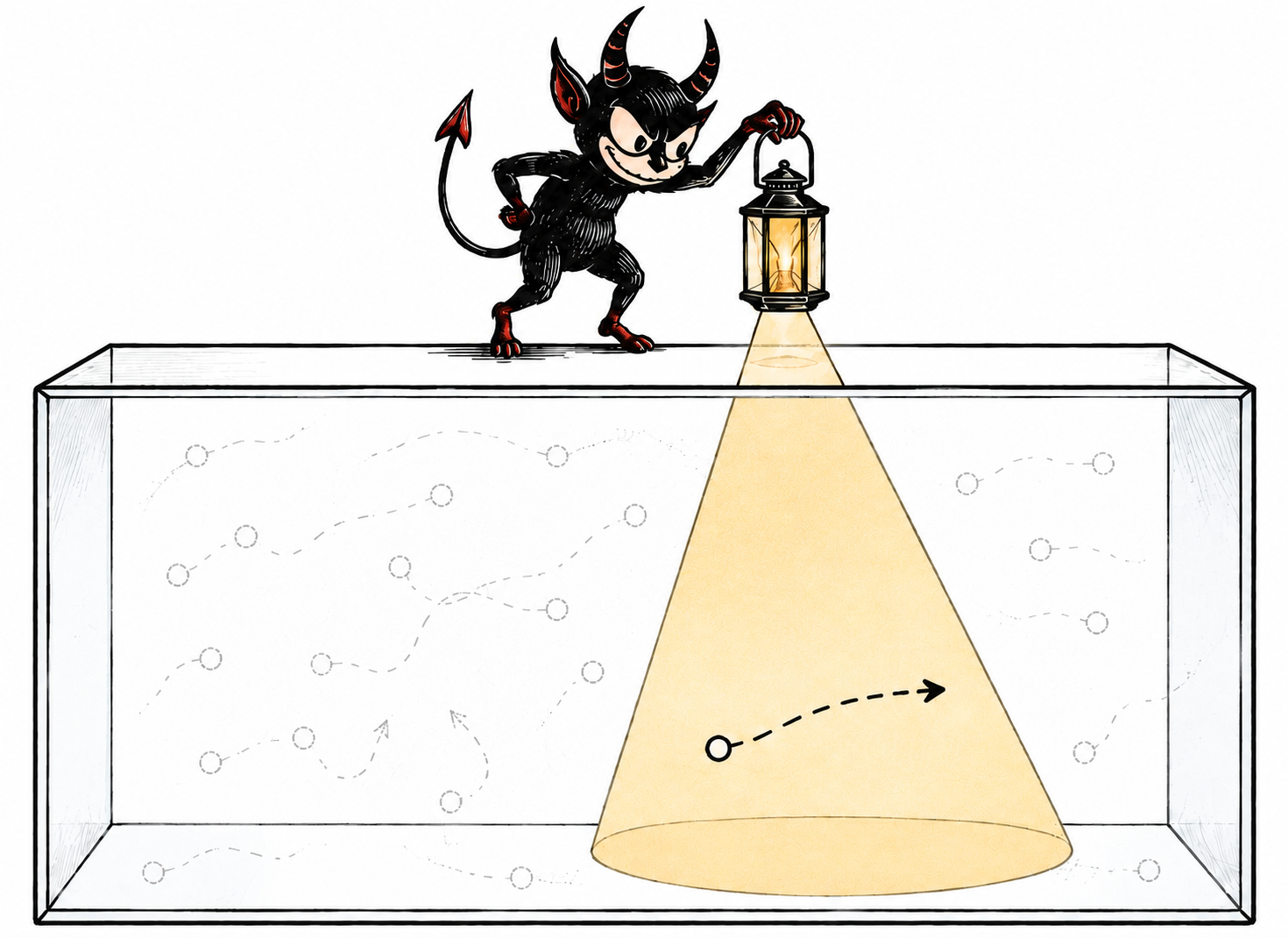}
    \caption{A Maxwell’s demon shines its lantern on particles moving in the dark to gather information about their positions and velocities.
}
    \label{fig: 1}
\end{figure}
\begin{equation}
  \left\langle e^{-S_{\rm tot}-I_c}\right\rangle = 1 ,\qquad \langle S_{\rm tot}\rangle \geq -\langle I_c\rangle .
  \label{eq:standard_feedback_ft}
\end{equation}
where $S_{\rm tot}$ denotes the total entropy production of the controlled system, $I_c$ is the information acquired from the measurement record, and the second expression corresponds to the generalized second law, obtained by applying Jensen’s inequality to the first equation.

However, Eq.~\eqref{eq:standard_feedback_ft} hides an assumption already pointed out in Refs.~\cite{Horowitz10,SagawaUeda12}: it requires the demon's measurement error to cover the whole phase space of the system being observed. In other words, independently of the state of the system, the demon could observe every outcome with nonzero probability. Mathematically, this condition is fulfilled when the measurement channel is assumed to have full support, but it fails when the channel contains zeros. Such singular measurements are not exceptional. They include ideal, error-free measurements, but also threshold detectors~\cite{Vidrighin16}, finite-resolution devices~\cite{Paneru18}, and sensors that do not have access to some parts of the phase space~\cite{Loos26}. In all these cases, the backward experiment may generate trajectories that are incompatible with the forward measurement record. These backward trajectories have no forward counterpart and therefore constitute a source of absolute irreversibility~\cite{Murashita14}. For the case of error-free measurements this effect is known to generate ``unavailable information"~\cite{Ashida14}: a part of the acquired information that cannot be converted into work because the time-reversed feedback protocol may fail to reproduce the measured outcome. 

In this paper I show that this phenomenon is more general: it is not tied to error-free measurements, but to the structure of the region of phase space that remains accessible to the demon through its measurement, even in the presence of errors. I derive fluctuation relations valid for arbitrary measurements by introducing a generalized notion of unavailable information. The result is a unified framework for feedback fluctuation theorems, which recovers both the standard full-support case~\cite{SagawaUeda12} and the error-free case~\cite{Ashida14} as limiting cases.

\paragraph{Preliminaries.}

Lets start by considering an stochastic thermodynamic system in contact with one or multiple heat reservoirs and driven by a control parameter $\lambda$. The system trajectory over the time interval $[0,\tau]$ is denoted by $X=\{x(t)\}_{0\leq t\leq \tau}$, where $x(t)$ is the state of the system at a given time. Measurements are performed over the system at different times $t_k$ producing the record $Y=\{y(t_1),\ldots,y(t_M)\}$, with $M$ being the number of measurements.  

The protocol applied after each measurement depends causally on the previous measurement record and is denoted by $\Lambda(Y)$. If we consider a fixed measurement record $\hat Y$, the  probability of having a given trajectory $X$ under the fixed protocol $\Lambda (\hat Y)$ is $P\bigl(X|\Lambda(\hat Y)\bigr)$. On the other hand, we can consider the probability of realizing the time-reversed version of the same trajectory $X^\dagger$ when applying the time reversed version of the protocol $\Lambda^\dagger(\hat Y)$ while no feedback control is performed over the system $P^{\dagger}\bigl(X^\dagger|\Lambda^\dagger(\hat Y)\bigr)$. In this situation, even thought that the forward dynamics occurs under the presence of feedback control we still have that the detailed fluctuation theorem holds 
\begin{equation}
  \frac{
  P\bigl(X|\Lambda(\hat Y)\bigr)
  }{
  P^{\dagger}\bigl(X^\dagger|\Lambda^\dagger(\hat Y)\bigr)
  }
  =
  e^{S_{\rm tot}(X,\hat Y)} ,
  \label{eq:fixed_protocol_dft}
\end{equation}
where $S_{\rm tot}(X,\hat Y)$ is the total entropy production along the system trajectory. The fact that Eq.~\eqref{eq:fixed_protocol_dft} holds at the trajectory level for any possible fixed protocol is crucial for deriving fluctuation theorems under the presence of feedback control as we will see. 

It is time now to define a central quantity in the context of stochastic thermodynamics of information, that is, the trajectory transfer entropy 
\begin{equation}
  I_c(X,Y)
  =
  \ln
  \frac{P_{c}(Y|X)}{P(Y)} ,
  \label{eq:information}
\end{equation}
where $P_c(Y|X)$ is the probability of measuring the outcome record $Y$ given a feedback-controlled trajectory $X$ and $P(Y)$ is the probability of the measurement record. This quantity can be interpreted as an stochastic analog of the mutual information between the system trajectory and the measurement record.

\begin{figure}
    \centering
    \includegraphics[width=0.8\linewidth]{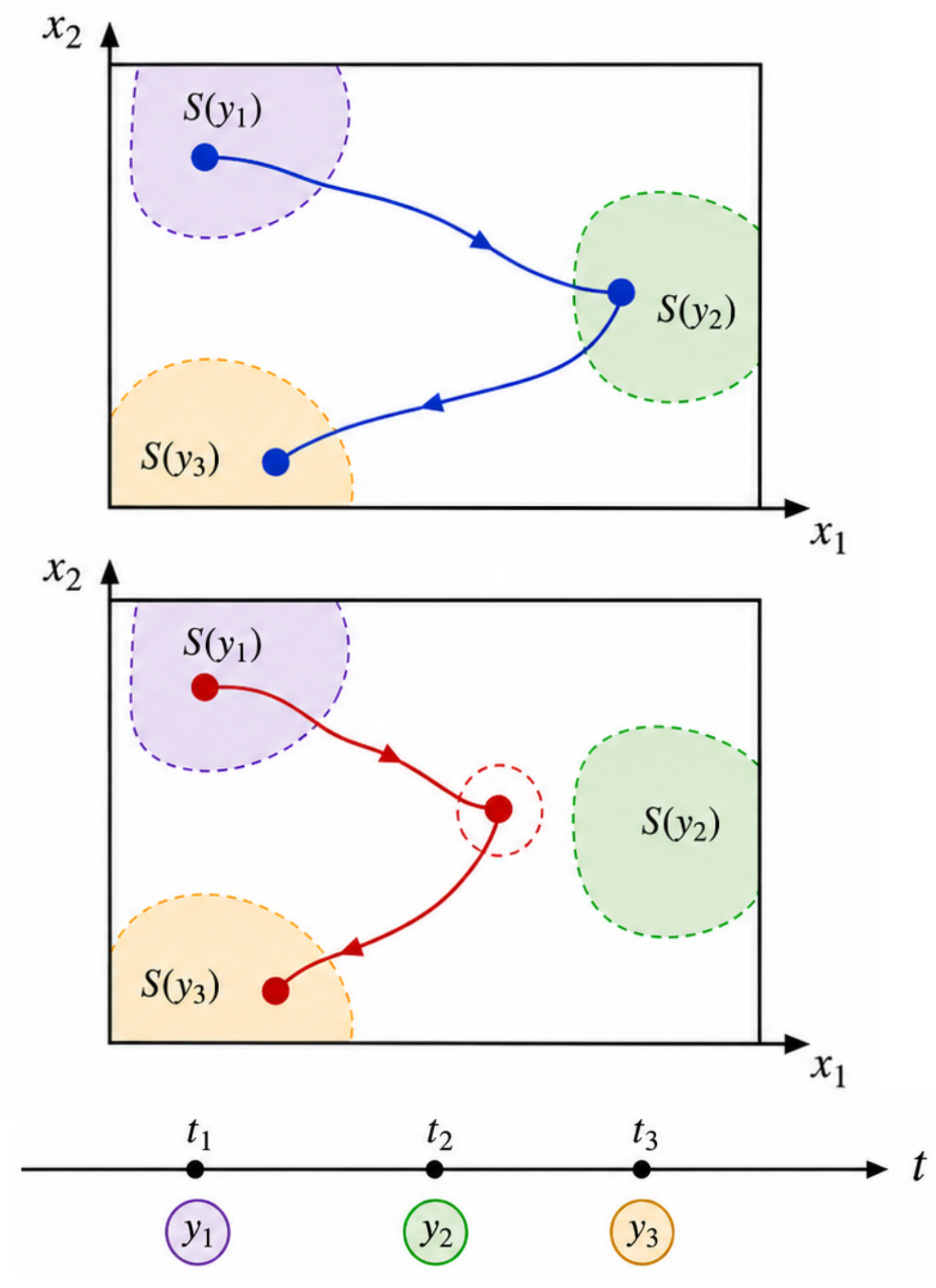}
    \caption{Examples of trajectories that are compatible (blue) and incompatible (red) with a given measurement record. The colored regions denote the parts of phase space associated with the outcomes obtained during the forward feedback-controlled process. In the backward experiment, the system evolves under the time-reversed protocol but without feedback, so its intrinsic stochastic dynamics may generate trajectories that leave these allowed regions. Such trajectories are incompatible with the original measurement record, as illustrated by the red path.}
    \label{fig: 2}
\end{figure}

When $P_{c}(Y|X)>0$ for every trajectory pair $(X,Y)$, Eqs.~\eqref{eq:fixed_protocol_dft}--\eqref{eq:information} directly recover the standard feedback fluctuation theorem, Eq.~\eqref{eq:standard_feedback_ft}, derived in Refs.~\cite{Horowitz10,SagawaUeda12}. We now relax this full-support assumption. For each measurement outcome $y$, we define $S(y)=\{x:P_{c}(y|x)>0\}$ as the region of phase space from which $y$ can occur with nonzero probability. Accordingly, a system trajectory $X$ is compatible with a measurement record $Y$ if $x(t_k)\in S(y(t_k))$ at every measurement time $t_k$ (see Fig.~\ref{fig: 2}). This condition is encoded by the indicator function $\chi_Y(X)$, with $\chi_Y(X)=1$ when $X$ is compatible with $Y$ and $\chi_Y(X)=0$ otherwise.

We require the backward dynamics generated by $\Lambda^\dagger(Y)$ to remain within the allowed region of phase space. The probability of this event is
\begin{equation}
  P_{\rm R}(Y)
  =
  \sum_X
  P^{\dagger}\bigl(X^\dagger|\Lambda^\dagger(Y)\bigr)
  \chi_Y(X^\dagger), 
  \label{eq:PR}
\end{equation}
note that generally $P_R(Y)<1$, since the stochastic backward dynamics without feedback can bring the system to not allowed parts of the phase-space. 

I define the generalized unavailable information as
\begin{equation}
  I_{\rm u}(Y)=-\ln P_{\rm R}(Y),
  \label{eq:Iu}
\end{equation}
which only depends on the measurement record and on the feedback protocol selected by that record. It vanishes when the measurement channel has full support, since then $\chi_Y(X^\dagger)=1$ for all trajectories and $P_{\rm R}(Y)=1$. For singular measurements, $I_{\rm u}$ quantifies the information acquired in the forward experiment that cannot be used reversibly by the chosen feedback protocol.

\paragraph{Fluctuation relation.}

We now turn to the derivation of our fluctuation relation. I keep the definition of the standard forward probability under feedback control used in Ref.~\cite{SagawaUeda12} 
\begin{equation}
    P(X,Y) = P_c(Y|X)P\bigl(X|\Lambda(Y)\bigr),
    \label{eq:forward}
\end{equation}
and to state the theorem, introduce the normalized backward distribution
\begin{equation}
  \widetilde P^\dagger(X^\dagger,Y)
  =
  \frac{
  P(Y)\,
  P^\dagger\bigl(X^\dagger|\Lambda^\dagger(Y)\bigr)
  \chi_Y(X^\dagger)
  }{
  P_{\rm R}(Y)
  },
  \label{eq:post_selected_backward}
\end{equation}
where the two distributions have well-defined operational meanings. The forward distribution represents the probability of having a given trajectory $X$ and a set of measurements $Y$ given that we are using such information at a given time to do some feedback on the system. On the other side, the backward distribution describes the probability of having the time-reversed trajectory $X^\dagger$ under the time-reversed fixed driving protocol $\Lambda ^\dagger (Y)$ and remaining on the allowed region of phase space. 

Using Eqs. (\ref{eq:information}), (\ref{eq:Iu}), (\ref{eq:forward}), and (\ref{eq:post_selected_backward}), we arrive at the main result of this paper in the form of the following generalized fluctuation relation
\begin{equation}
  \frac{
  P(X,Y)
  }{
  \widetilde P^\dagger(X^\dagger,Y)
  }
  =
  e^{
  S_{\rm tot}(X,Y)+I_c(X,Y)-I_{\rm u}(Y)
 },
  \label{eq:main_detailed}
\end{equation}
which is valid for arbitrary feedback and measurement channels.  
Averaging over forward realizations gives the integral fluctuation relation (see Appendix~\ref{app: ft} for details)
\begin{equation}
  \left\langle
  e^{-S_{\rm tot}-I+I_{\rm u}}
  \right\rangle
  =1 .
  \label{eq:main_ift}
\end{equation}

The results in Eqs.~\eqref{eq:main_detailed} and~\eqref{eq:main_ift} provide a unified framework for feedback-controlled systems, containing as limiting cases the two apriori qualitatively different formulations presented in Refs.~\cite{SagawaUeda12,Ashida14}. In the full-support limit, $P_{c}(Y|X)>0$ for all $(X,Y)$, so that $\chi_Y(X^\dagger)=1$ and $P_{\rm R}(Y)=1$. Consequently, $I_{\rm u}=0$, and Eq.~\eqref{eq:main_ift} reduces to Eq.~\eqref{eq:standard_feedback_ft}. The error-free limit of Ref.~\cite{Ashida14} is recovered when the measurement record is a deterministic function of the trajectory, $Y=f(X)$. In this case, the support of the measurement channel contains only trajectories that produce the same record, and $P_{\rm R}(Y)$ is the probability that the backward protocol reproduces the forward measurement record. Consequently, $I_{\rm u}$ coincides with the unavailable information introduced for error-free measurements. 

Applying Jensen's inequality to Eq.~\eqref{eq:main_ift} we can get the generalized second law,
\begin{equation}
  \langle S_{\rm tot}\rangle
  \geq
  -\langle I_c\rangle+\langle I_{\rm u}\rangle \geq -\langle I_c \rangle , 
  \label{eq:second_law}
\end{equation}
where the last inequality holds since $\langle I_{\rm u}\rangle \geq 0$ by definition. Thus, the thermodynamic advantage provided by feedback is bounded not by the acquired information alone, but by the available information, $\Delta I = I - I_{\rm u}$. The unavailable contribution consequently tightens the usual information-theoretic second law [Eq.~\eqref{eq:standard_feedback_ft}].

In addition to the generalized integral fluctuation theorem in Eq.~\eqref{eq:main_ift}, two further integral relations follow from the detailed fluctuation relation in Eq.~\eqref{eq:main_detailed} (see Appendix~\ref{app: ft} for details). The first reads
\begin{equation}\label{eq: efficacy 1}
\left\langle e^{-S_{\rm tot}-I_c}\right\rangle
=
\left\langle e^{-I_u}\right\rangle,
\end{equation}
while the second is given by
\begin{equation}\label{eq: efficacy 2}
\left\langle e^{-S_{\rm tot}}\right\rangle
=
\left\langle e^{I_c-I_u}\right\rangle_{\tilde{P}^\dagger},
\end{equation}
where the right-hand side of Eq.~\eqref{eq: efficacy 2} can be identified as a generalization of the efficacy parameter $\gamma$ introduced in Ref.~\cite{SagawaUeda12}. Moreover, Eqs.~\eqref{eq: efficacy 1} and~\eqref{eq: efficacy 2} generalize the fluctuation theorems derived in Ref.~\cite{Archambault25} for the particular case of error-free measurements.

\paragraph{Maxwell demon with finite resolution.}

\begin{figure}[b!]
    \centering
    \includegraphics[width=1.05\linewidth]{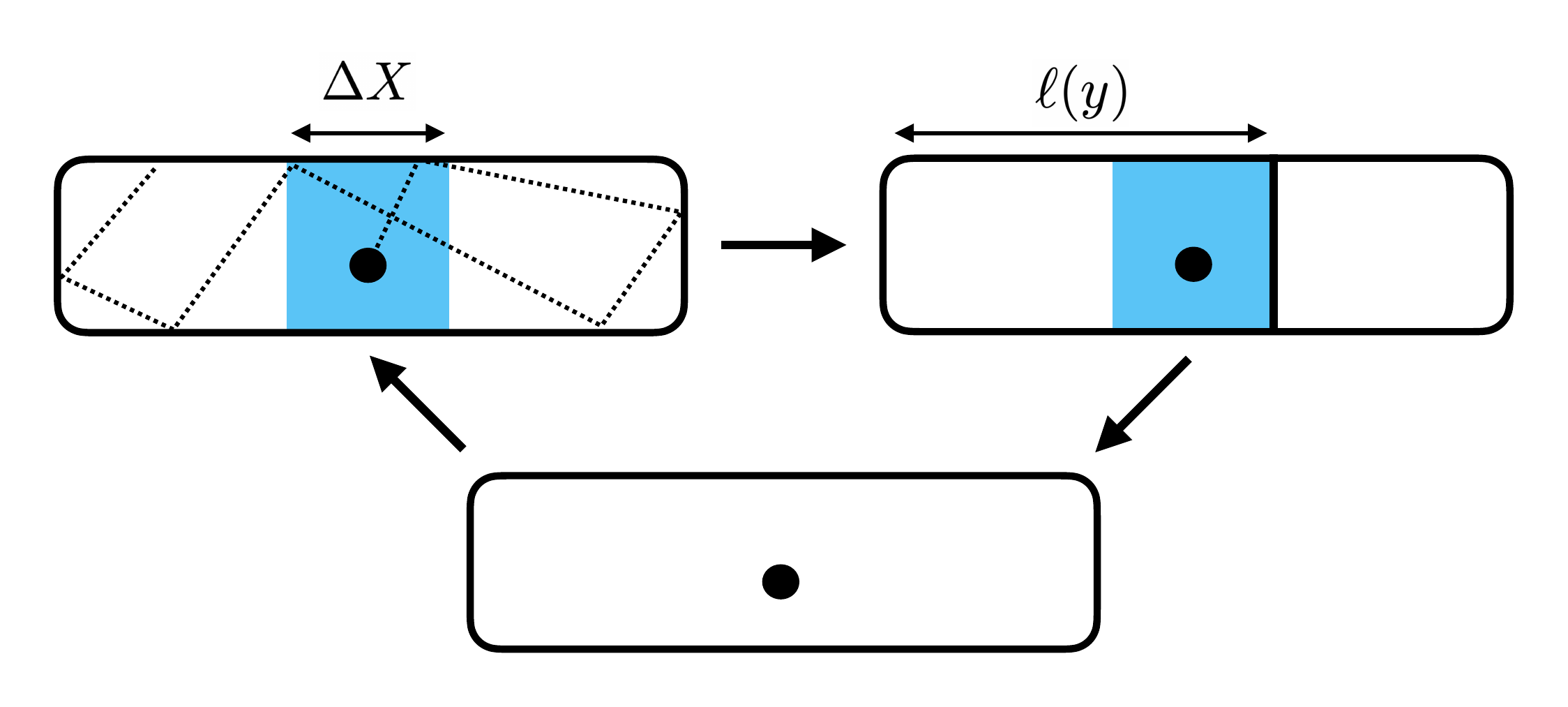}
    \caption{Szilard engine with spatial resolution $\Delta X$. The demon localizes the particle to within one pixel, so decreasing the pixel size allows it to extract more information.
}
    \label{fig: 3}
\end{figure}

We now illustrate the validity of the main theorem with a single-particle Szilard engine (see Fig.~\ref{fig: 3}). A Brownian
particle is confined in a one-dimensional box of length $L$ and is initially
equilibrated, so that $p_0(x)=1/L$. The demon measures the position with finite
resolution $\Delta X$ for a given outcome $y$, the compatible region of phase-space is $S(y)=[a(y),b(y)]$, with $a(y)=\max\left(0,y-\Delta X/2\right)$ and $b(y)=\min\left(L,y+\Delta X/2\right)$.
Let $\delta(y)=b(y)-a(y)$ be the length of this interval. In the bulk of the box, one simply has \(\delta(y)=\Delta X\). Close to the boundaries, however, the compatible interval is truncated by the hard walls and \(\delta(y)<\Delta X\). The probability of the measurement outcomes would be
\begin{equation}
  P(y)=
    \int_0^L dx\, p_0(x)P_c(y|x)=\frac{\delta(y)}{L\Delta X},
\end{equation}
together with the acquired information associated with the measurement
\begin{equation}
  I(x,y)=\ln\frac{L}{\delta(y)},\qquad x \in S(y). 
  \label{eq:szilard_info}
\end{equation}
where if $\delta(y)=\Delta X$, we obtains $I(y)=\ln (L/\Delta X)$. But near the hard walls, the compatible interval $\delta (y)$ is smaller and the measurement outcome is correspondingly more informative for the demon.

The demon now performs a feedback operation using a movable wall. Since the particle is only known to lie somewhere inside $S(y)$, the wall can be inserted safely only at one of the two endpoints of that interval. The optimal one-wall protocol chooses the smaller initial expansion
length, $\ell(y)=\min\{b(y),L-a(y)\}$. The particle is then expanded quasi-statically and isothermally to the full box. The extracted work after the expansion is
\begin{equation}
  \beta W_{\rm ext}(y)=\ln\frac{L}{\ell(y)} ,
  \label{eq:work_ext}
\end{equation}
and because the demon is performing a cycle the total entropy production of the controlled system would be
\begin{equation}
  S_{\rm tot}(y)=-\beta W_{\rm ext}(y).
\end{equation}

In the backward experiment, the selected protocol compresses the particle
quasi-statically from the full box to the interval of length $\ell(y)$. At the
time conjugate to the measurement, the particle is uniformly distributed over
this interval. The probability that it lies inside the forward-compatible
phase-space $S(y)$ is therefore
\begin{equation}
  P_{\rm R}(y)=\frac{\delta(y)}{\ell(y)} \implies
  I_{\rm u}(y)=\ln\frac{\ell(y)}{\delta(y)},
  \label{eq:szilard_Iu}
\end{equation}
we can see that near the walls, the compatible interval $S(y)$ can be isolated by placing the piston wall at its inner endpoint, because the other endpoint is already provided by the hard wall. In that case \(\ell(y)=\delta(y)\), and hence $I_{\rm u} =0$. The information acquired from such boundary outcomes is fully available for work extraction by the demon. More generally, the available information for the demon is
\begin{equation}
  \Delta I(y)
  =
  \ln\frac{L}{\ell(y)}
  =
  \beta W_{\rm ext}(y).
  \label{eq:available_equals_work}
\end{equation}
Thus the generalized fluctuation theorem is saturated for every measurement
\begin{equation}
  S_{\rm tot}(y)+\Delta I(y)=0.
\label{eq:brach_SL}
\end{equation}

We can also compute the averaged quantities explicitly. Let $r=\Delta X/L$ (see Appendix~\ref{app:explicit_averages} for details). Using Eq.~\eqref{eq:szilard_info}, the average acquired information is
\begin{equation}
    \langle I\rangle
    =
    \int dy\, P(y)\ln\frac{L}{\delta(y)}
    =
    -\ln r
    +
    \frac{r}{2},
    \label{eq:avg_Ic}
\end{equation}
where the second term comes from the boundary regions in which the compatible interval is truncated by the hard walls.

The average extracted work using Eq.~\eqref{eq:work_ext} is
\begin{equation}
    \beta\langle W_{\rm ext}\rangle
    =
    (1+r)\ln\frac{2}{1+r}
    +
    1-\frac{r}{2}
    -
    r\ln\frac{1}{r}.
    \label{eq:avg_work}
\end{equation}
Similarly, the average unavailable information is
\begin{equation}
    \langle I_u\rangle
    =
    \int dy\, P(y)\ln\frac{\ell(y)}{\delta(y)}
    =
    (1+r)\ln\frac{1+r}{2r}
    +
    r-1.
    \label{eq:avg_Iu}
\end{equation}
Equations~\eqref{eq:avg_Ic}--\eqref{eq:avg_Iu} satisfy
\begin{equation}
    \beta\langle W_{\rm ext}\rangle
    =
    \langle I_c\rangle-\langle I_u\rangle,
\end{equation}
as expected from the identity in Eq.~\eqref{eq:brach_SL}.

In the high-resolution limit $\Delta X/L\to 0$, the acquired information by the demon grows logarithmically,
\begin{equation}
  \langle I\rangle\simeq \ln\frac{L}{\Delta X},
\end{equation}
meaning that, as the measurement resolution becomes arbitrarily fine, the demon localizes the particle with a increasingly high precision, and the information obtained by the measurement diverges. 

However, the unavailable information diverges with exactly the same leading contribution,
\begin{equation}
  \langle I_{\rm u}\rangle
  \simeq
  \ln\frac{L}{\Delta X}-1-\ln 2.
\end{equation}
Their difference remains finite and equal to the extractable work 
\begin{equation}
  \beta\langle W_{\rm ext}\rangle
  \simeq
  1+\ln 2, 
\end{equation}
showing that arbitrarily precise measurements do not necessarily lead to arbitrary large work extraction. Although the demon acquires an unbounded amount of information about the particle position, most of this additional information becomes unavailable to the permitted feedback protocol. The limitation is therefore not the accuracy of the measurement record, but the restricted set of operations through which that information can be exploited. Only the finite difference between the acquire and unavailable information is operationally useful for extracting work. 

\paragraph{Conclusions.}

I have derived a fluctuation relation for feedback-controlled systems that unifies, within a single framework, the existing informational formulations of nonequilibrium feedback control. In the intermediate regime between the two limiting formulations it connects, the relation applies to a broad class of scenarios inaccessible to any previous formulation. The Szilard-engine example illustrates one of such scenarios, where the unavailable information can arise not only from limited measurement resolution itself, but from the physical limitations of the feedback operation. Such constraints are expected to be central in realistic feedback-control settings, including biological Maxwell demons, whose feedback mechanisms are physically restricted rather than arbitrary and for which unavailable information may prove essential to understanding the thermodynamic limitations of biological systems that use information for functional purposes. A second scenario of interest is continuous feedback control, where the information acquired through continuous measurement may diverge. The present framework offers a route to determine whether an equally divergent unavailable contribution accounts for why this information cannot be converted into extractable work. Finally, it will be interesting to compare the bounds derived here with non-informational bounds on entropy production in feedback-controlled systems, such as those based on coarse-grained entropy production~\cite{Potts18}.

\section{acknowledgments}

I thank Gonzalo Manzano and Juan M. R. Parrondo for useful discussions. I acknowledge support from the CoQuSy project PID2022-140506NB-C21 and the Mar\'ia de Maeztu project CEX2021-001164-M for Units of Excellence, funded by MICIU/AEI/10.13039/501100011033/FEDER, UE, and Conselleria d'Educaci\'o, Universitat i Recerca of the Balearic Islands through Grant FPI 058 2022.

\bibliographystyle{apsrev4-2}

\bibliography{refs}
\newpage
\onecolumngrid

\setcounter{secnumdepth}{1}
\appendix

\section{Proofs of the main fluctuation relations}
\label{app: ft}
In this Appendix, we explicitly derive the main integral fluctuation relations presented in the main text. 

To obtain Eq.~\eqref{eq:main_ift}, we average Eq.~\eqref{eq:main_detailed} over the forward realizations
\[
\begin{split}
    &\langle e^{-S_{\rm{tot}}-I_c+I_u} \rangle = \sum_{X}\sum_{Y} P(X,Y) e^{-S_{\rm{tot}}-I_c+I_u} = \sum_{X}\sum_{Y} P(X,Y) \dfrac{\tilde P^\dagger(X^\dagger, Y)}{P(X,Y)} = \sum_{Y}\dfrac{P(Y)}{P_R(Y)}\sum_{X}
  P^\dagger\bigl(X^\dagger|\Lambda^\dagger(Y)\bigr)
  \chi_Y(X^\dagger) \\& = \sum_Y \dfrac{P(Y)}{P_R(Y)} P_R(Y) = \sum_Y P(Y) = 1,
\end{split}
\]
where for the penultimate step, we used the definition of $P_R(Y)$ to evaluate the sum over $X$, while the final equality follows from the normalization of the measurement-record distribution $P(Y)$. 

Equation~\eqref{eq: efficacy 1} also follows directly from the detailed fluctuation relation, Eq.~\eqref{eq:main_detailed}. Averaging over the forward realizations, we obtain
\[
    \begin{split}
        &\left \langle e^{-S_{\rm tot}-I_c} \right\rangle = \sum_X\sum_Y P(X,Y)  e^{-S_{\rm tot}-I_c} = \sum_X\sum_Y P(X,Y) \dfrac{P^\dagger (X^\dagger|\Lambda^\dagger(Y))\chi_Y(X^\dagger)P(Y)}{P_c(Y|X)P(X|\Lambda(Y))} = \sum_X\sum_Y P^\dagger (X^\dagger|\Lambda^\dagger(Y))\chi_Y(X^\dagger)P(Y)\\& = \sum_Y P(Y) e^{-I_{\rm u}} = \langle e^{-I_{\rm u}} \rangle, 
    \end{split}
\]
where the sum over $X$ was evaluated using the definition of $P_R(Y)$, together with $P_R(Y)=e^{-I_u(Y)}$.

Finally, Eq.~\eqref{eq: efficacy 2} can be derived from the detailed fluctuation relation, Eq.~\eqref{eq:main_detailed}, by following a similar procedure
\[
\begin{split}
&\langle e^{-S_{\rm tot}} \rangle = \sum_X\sum_Y P(X,Y)e^{-S_{\rm tot}} = \sum_X\sum_Y P(X,Y) \dfrac{P^\dagger(X^\dagger|\Lambda^\dagger(Y))\chi_Y(X^\dagger)}{P(X|\Lambda(Y))} = \sum_X \sum_Y P_c(Y|X) P^\dagger(X^\dagger|\Lambda^\dagger(Y))\chi_Y(X^\dagger)\\& = \sum_X \sum_Y \dfrac{P^\dagger(X^\dagger|\Lambda^\dagger(Y))\chi_Y(X^\dagger)P(Y)}{P_R(Y)}\dfrac{P_c(Y|X)}{P(Y)}P_R(Y)= \sum_X\sum_Y \tilde P^\dagger(X^\dagger,Y)e^{I_c - I_{\rm u}} = \langle e^{I_c - I_{\rm u}} \rangle_{\tilde P^\dagger}, 
\end{split} 
\]
where for arriving to the last expression we just multiply and divide by $P(Y)$ and $P_R(Y)$. 

\section{Explicit calculation of the averaged quantities for the Szilard engine example}
\label{app:explicit_averages}

In this Appendix, we explicitly calculate the average quantities for the finite-resolution Szilard engine discussed in the main text. We consider a particle confined to a one-dimensional box of length $L$, with initial equilibrium distribution
\begin{equation}
    p_0(x)=\frac{1}{L},
    \qquad
    0\leq x\leq L.
\end{equation}

The position measurement has spatial resolution $\Delta X$. For a measurement outcome $y$, the set of positions compatible with
that outcome is
\begin{equation}
    S(y)=[a(y),b(y)],
\end{equation}
where $a(y)=\max\left(0,y-\frac{\Delta X}{2}\right)$ and $b(y)=\min\left(L,y+\frac{\Delta X}{2}\right)$. The length of the compatible interval is therefore $\delta(y)=b(y)-a(y)$. 
Explicitly,
\begin{equation}
\delta(y)=
\begin{cases}
    y+\dfrac{\Delta X}{2},
    &
    -\dfrac{\Delta X}{2}
    \leq y\leq
    \dfrac{\Delta X}{2},
    \\[3mm]
    \Delta X,
    &
    \dfrac{\Delta X}{2}
    \leq y\leq
    L-\dfrac{\Delta X}{2},
    \\[3mm]
    L-y+\dfrac{\Delta X}{2},
    &
    L-\dfrac{\Delta X}{2}
    \leq y\leq
    L+\dfrac{\Delta X}{2}.
\end{cases}
\label{eq:app_delta_piecewise}
\end{equation}

The probability density of obtaining the outcome $y$ is
\begin{equation}
    P(y)
    =
    \int_0^L dx\,p_0(x)P_c(y|x)
    =
    \frac{\delta(y)}{L\Delta X}.
    \label{eq:app_py}
\end{equation}

\subsection{Average acquired information}

The stochastic information acquired from the measurement is
\begin{equation}
    I_c(x,y)
    =
    \ln\frac{P_c(y|x)}{P(y)}
    =
    \ln\frac{L}{\delta(y)},
    \qquad
    x\in S(y).
    \label{eq:app_information}
\end{equation}
for this measurement, the information depends on the outcome
$y$ but not explicitly on the position $x$ within $S(y)$. Its average
is therefore
\begin{equation}
    \langle I_c\rangle
    =
    \int_{-\Delta X/2}^{L+\Delta X/2}
    dy\,P(y)\ln\frac{L}{\delta(y)}.
    \label{eq:app_I_average_definition}
\end{equation}

Using Eq.~\eqref{eq:app_delta_piecewise}, we separate the two boundary regions from the bulk
\begin{align}
\langle I_c\rangle
={}&
\int_{-\Delta X/2}^{\Delta X/2}
dy\,
\frac{y+\Delta X/2}{L\Delta X}
\ln\frac{L}{y+\Delta X/2}
\nonumber+
\int_{\Delta X/2}^{L-\Delta X/2}
\frac{dy}{L}
\ln\frac{L}{\Delta X}
\nonumber+
\int_{L-\Delta X/2}^{L+\Delta X/2}
dy\,
\frac{L-y+\Delta X/2}{L\Delta X}
\ln\frac{L}{L-y+\Delta X/2}.
\label{eq:app_I_split}
\end{align}

The two boundary integrals are equal by reflection symmetry. Hence,
\begin{align}
\langle I_c\rangle
=
\frac{2}{L\Delta X}
\int_{-\Delta X/2}^{\Delta X/2}
dy\,
\left(y+\frac{\Delta X}{2}\right)
\ln\frac{L}{y+\Delta X/2}
\nonumber+
\left(1-\frac{\Delta X}{L}\right)
\ln\frac{L}{\Delta X},
\end{align}
by introducing $u=y+\Delta X/2$, the boundary integral becomes $2/(L\Delta X)\int_0^{\Delta X}  du\,u\ln (L/u)$. Using
\begin{equation}
    \int du\,u\ln\frac{L}{u}
    =
    \frac{u^2}{2}\ln\frac{L}{u}
    +\frac{u^2}{4},
    \label{eq:app_integral_u_log}
\end{equation}
together with $\lim_{u\rightarrow0^+}u^2\ln u=0$, we obtain
\begin{align}
    \frac{2}{L\Delta X}
    \int_0^{\Delta X} du\,
    u\ln\frac{L}{u}
    =
    \frac{2}{L\Delta X}
    \left[
    \frac{u^2}{2}\ln\frac{L}{u}
    +\frac{u^2}{4}
    \right]_0^{\Delta X}
    \nonumber=
    \frac{\Delta X}{L}
    \left(
    \ln\frac{L}{\Delta X}
    +\frac{1}{2}
    \right).
    \label{eq:app_I_boundary_result}
\end{align}

Consequently,
\begin{align}
    \langle I_c\rangle
    =
    \frac{\Delta X}{L}
    \left(
    \ln\frac{L}{\Delta X}
    +\frac{1}{2}
    \right)
    +
    \left(1-\frac{\Delta X}{L}\right)
    \ln\frac{L}{\Delta X}
    \nonumber
    =
    \ln\frac{L}{\Delta X}
    +\frac{\Delta X}{2L}.
\end{align}

In terms of $r=\Delta X/L$, the average acquired information is
\begin{equation}
    \langle I_c\rangle
    =
    -\ln r+\frac{r}{2}.
    \label{eq:app_I_final}
\end{equation}

\subsection{Average extracted work}

For a given outcome $y$, the movable wall can be safely inserted at one of the endpoints of the compatible interval $S(y)$. The optimal one-wall protocol selects the smaller expansion length,
\begin{equation}
    \ell(y)
    =
    \min\left\{
    b(y),L-a(y)
    \right\}.
    \label{eq:app_ell_definition}
\end{equation}

Its explicit form is
\begin{equation}
\ell(y)=
\begin{cases}
    y+\dfrac{\Delta X}{2},
    &
    -\dfrac{\Delta X}{2}
    \leq y\leq
    \dfrac{L}{2},
    \\[3mm]
    L-y+\dfrac{\Delta X}{2},
    &
    \dfrac{L}{2}
    \leq y\leq
    L+\dfrac{\Delta X}{2}.
\end{cases}
\label{eq:app_ell_piecewise}
\end{equation}

The work extracted during the quasistatic isothermal expansion is
\begin{equation}
    \beta W_{\mathrm{ext}}(y)
    =
    \ln\frac{L}{\ell(y)},
    \label{eq:app_work_branch}
\end{equation}
and the average extracted work would be
\begin{equation}
    \beta\langle W_{\mathrm{ext}}\rangle
    =
    \int_{-\Delta X/2}^{L+\Delta X/2}
    dy\,P(y)\ln\frac{L}{\ell(y)}.
    \label{eq:app_work_average_definition}
\end{equation}

Using reflection symmetry around $y=L/2$, we may write
\begin{align}
\beta\langle W_{\mathrm{ext}}\rangle
=
2\int_{-\Delta X/2}^{\Delta X/2}
dy\,
\frac{y+\Delta X/2}{L\Delta X}
\ln\frac{L}{y+\Delta X/2}
\nonumber+
2\int_{\Delta X/2}^{L/2}
\frac{dy}{L}
\ln\frac{L}{y+\Delta X/2}.
\label{eq:app_work_split}
\end{align}

The first term is the total contribution from the two boundary regions
\begin{equation}
    \langle W_{\mathrm{boundary}} \rangle 
    =
    \frac{\Delta X}{L}
    \left(
    \ln\frac{L}{\Delta X}
    +\frac{1}{2}
    \right)
    =
    r\left(
    \ln\frac{1}{r}
    +\frac{1}{2}
    \right).
    \label{eq:app_work_boundary}
\end{equation}

For the bulk contribution, we introduce $u=y+\Delta X/2$. The integration limits are then $u=\Delta X$ and $u=(L+\Delta X)/2$, so that
\begin{equation}
    \langle W_{\mathrm{bulk}}\rangle
    =
    \frac{2}{L}
    \int_{\Delta X}^{(L+\Delta X)/2}
    du\,\ln\frac{L}{u}.
    \label{eq:app_work_bulk_integral}
\end{equation}

Using $\int du\,\ln(L/u)=u\ln(L/u)+u$, we find
\begin{align}
\langle W_{\mathrm{bulk}} \rangle
=
\frac{2}{L}
\left[
u\ln\frac{L}{u}+u
\right]_{\Delta X}^{(L+\Delta X)/2}
\nonumber&=
\frac{2}{L}
\left[
\frac{L+\Delta X}{2}
\ln\frac{2L}{L+\Delta X}
+
\frac{L+\Delta X}{2}
-
\Delta X\ln\frac{L}{\Delta X}
-
\Delta X
\right]
\nonumber\\&=
(1+r)\ln\frac{2}{1+r}
+1-r
-2r\ln\frac{1}{r}.
\label{eq:app_work_bulk_result}
\end{align}

Adding the boundary and bulk contributions gives
\begin{align}
\beta\langle W_{\mathrm{ext}}\rangle
=
r\left(
\ln\frac{1}{r}
+\frac{1}{2}
\right)
+
(1+r)\ln\frac{2}{1+r}
\nonumber+
1-r
-2r\ln\frac{1}{r} = (1+r)\ln\frac{2}{1+r}
    +1-\frac{r}{2}
    -r\ln\frac{1}{r}.
\end{align}

\subsection{Average unavailable information}
In the backward experiment, the particle is uniformly distributed
over an interval of length $\ell(y)$. The probability of finding it
inside the forward-compatible interval is
\begin{equation}
    P_R(y)
    =
    \frac{\delta(y)}{\ell(y)},
\end{equation}
and the unavailable information is therefore
\begin{equation}
    I_u(y)
    =
    -\ln P_R(y)
    =
    \ln\frac{\ell(y)}{\delta(y)}.
    \label{eq:app_Iu_branch}
\end{equation}

In the boundary regions, one has $\ell(y)=\delta(y)$, and hence $I_u(y)=0$. Only the bulk outcomes contribute to the average unavailable information. Using left--right symmetry,
\begin{equation}
    \langle I_u\rangle
    =
    2\int_{\Delta X/2}^{L/2}
    \frac{dy}{L}
    \ln\frac{y+\Delta X/2}{\Delta X},
    \label{eq:app_Iu_integral_y}
\end{equation}
with $u=y+\Delta X/2$, this becomes
\begin{equation}
    \langle I_u\rangle
    =
    \frac{2}{L}
    \int_{\Delta X}^{(L+\Delta X)/2}
    du\,\ln\frac{u}{\Delta X}.
    \label{eq:app_Iu_integral_u}
\end{equation}

Using $\int du\,\ln(u/\Delta X)=u\ln(u/\Delta X)-u$, we obtain
\begin{align}
\langle I_u\rangle
=
\frac{2}{L}
\left[
u\ln\frac{u}{\Delta X}-u
\right]_{\Delta X}^{(L+\Delta X)/2}
\nonumber=
\frac{2}{L}
\left[
\frac{L+\Delta X}{2}
\ln\frac{L+\Delta X}{2\Delta X}
-
\frac{L+\Delta X}{2}
+
\Delta X
\right].
\end{align}

Therefore,
\begin{equation}
    \langle I_u\rangle
    =
    (1+r)\ln\frac{1+r}{2r}
    +r-1.
    \label{eq:app_Iu_final}
\end{equation}

\subsection{High-resolution limit}

We finally consider the high-resolution limit $r\rightarrow0$. From Eq.~\eqref{eq:app_I_final},
\begin{equation}
    \lim_{r\rightarrow0} \langle I_c\rangle
    = \lim_{r\rightarrow0}(
    -\ln r+\frac{r}{2})
    =
    \ln\frac{L}{\Delta X}.
    \label{eq:app_I_asymptotic}
\end{equation}

Thus, the acquired information diverges logarithmically as the measurement resolution increases. For the unavailable information,
\begin{align}
    \lim_{r\rightarrow0} \langle I_u\rangle
    = \lim_{r\rightarrow0}\left (
    (1+r)
    \left[
    \ln(1+r)-\ln 2-\ln r
    \right]
    +r-1\right)
    \nonumber
    =
    -\ln r-1-\ln 2.
\end{align}

The acquired and unavailable information contain the same leading logarithmic divergence. Their difference remains finite
\begin{align}
    \beta\langle W_{\mathrm{ext}}\rangle=
    \langle I_c\rangle-\langle I_u\rangle
    =
    1+\ln 2.
\end{align}

\end{document}